\documentclass[]{spie}  

\usepackage{amsmath,amsfonts,amssymb}
\usepackage{graphicx}
\usepackage[colorlinks=true, allcolors=blue]{hyperref}

\title{Laser-micromachined silicon-platelet feedhorns for large-scale submillimeter and millimeter-wave focal planes}

\author[a]{Jason E. Austermann}
\author[a]{James Beall}
\author[b]{Andrew Forsman}
\author[b]{Haibo Huang}
\author[a]{Johannes Hubmayr}
\author[a,c]{Matthew A. Koc}
\author[a]{Jeff van Lanen}
\author[b]{Pavel Lapa}
\author[b]{Josh Raimist}
\author[d]{Sara M. Simon}
\author[b]{Jordan Stutz}
\author[b]{Anatolios Tambazidis}
\author[a]{Joel N. Ullom}
\affil[a]{Quantum Sensors Division, National Institute of Standards and Technology, 325 Broadway, Boulder, USA}
\affil[b]{Inertial Fusion Technology Division, Energy Group, General Atomics, San Diego, USA}
\affil[c]{Department of Physics, University of Colorado, Boulder, CO 80309 USA}
\affil[d]{Fermi National Accelerator Laboratory, Batavia, IL 60510, USA}

\authorinfo{E-mail: jea at nist.gov}

\begin{document} 
\maketitle

\begin{abstract}
We present the fabrication and characterization of the first silicon-platelet feedhorn arrays produced using laser micromachining. 
First, we present a demonstration of the technology for the millimeter-wave band of 80~GHz to 170~GHz, i.e. covering the 90/150~GHz bands typical of CMB experiments. 
Next, we expand the technology to large-scale production on 150~mm wafers and demonstrate operation at submillimeter wavelengths. 
This feedhorn array is optimized for operation in a band centered at 350~GHz (330~GHz to 370~GHz) and is being deployed as one of the focal plane elements of the CCAT 350~GHz module of Prime-Cam. 
We present the design and fabrication processes for these feedhorn arrays and 
compare the optical performance directly to simulation and to feedhorns of identical design but produced using traditional deep reactive-ion etching (DRIE). 
We conclude with a discussion of future expansions of this technology, including the potential of sidewall control and using thicker (and thus fewer) wafers, which could significantly reduce production costs and labor.
\end{abstract}


\section{INTRODUCTION}
\label{sec:intro}  

Experiments operating at millimeter and submillimeter wavelengths continue to grow in scale to reach better sensitivities and higher mapping speeds.  
Projects like the Simons Observatory\cite{zhu2021simons,galitzki2024simons}, CCAT Prime-Cam\cite{vavagiakis2018ccatprime}, and the combined cosmic microwave background (CMB) observatories at the South Pole\cite{natoli2026_stp3gplus_temporary,hui2018bicep} are all building or operating detector counts near 100,000 and beyond. 
High precision experiments, like those measuring the polarization of the CMB, also require tight control of systematics\cite{abitbol2017cmb} as they aim to detect tiny fluctuations that are often buried in larger signals.
Feedhorns can be an excellent choice for systematic control at millimeter and sub-millimeter wavelengths due to their ability to produce well formed beams, low cross polarization, wide bandwidth, and their strong pixel-to-pixel isolation that results in low optical cross talk and strong stray light control.  

Silicon-platelet feedhorns\cite{Britton2010,nibarger201284,simon2016,austermann26} can provide particularly strong systematic control when produced using high precision manufacturing with high uniformity across large scales. 
Such micromachined silicon-substrate feedhorns have proven to be a robust and reliable solution for precise and uniform coupling of millimeter-wave radiation to cryogenic detectors for a number of experiments\cite{austermann2012sptpol,niemack2010actpol,henderson2016,austermann2018millimeter,hubmayr2016spider,vavagiakis2022ccat,austermann26}. 
Large arrays have traditionally been manufactured by stacking metalized silicon platelets, each with lithographically defined holes produced using deep reactive-ion etching (DRIE).  
These arrays have typically been built from $\sim$30 to 50 individual 150~mm diameter wafers – a large but achievable task for past experiments only requiring a few arrays. 
However, scaling the current production from a few arrays to several tens, or hundreds, would likely prove impractical for a research level cleanroom and experience little to no cost reduction at scale. 

Here we demonstrate an alternative approach: laser micromachining of silicon wafers.  
Both lithographic DRIE and laser machining\cite{wang2021review_silicon_machining} processes are capable of providing feature size and positioning precision of a few $\mu$m, or better, with large-scale uniformity.  This is significantly better than the $\sim$20~$\mu$m precision one might expect from high-precision, traditional direct machining of aluminum and other metals. 
This precision becomes increasingly important at higher frequencies (shorter wavelengths) of operation.
Here, we explore laser micromachining as a scalable solution that can take much of the production out of the cleanroom and provide potential advantages in fabrication speed, labor, throughput, and capability. 

We present the fabrication and characterization of the first 
silicon-platelet feedhorn arrays produced using laser micromachining.  
First, we present a mid-scale demonstration of laser-micromachined feedhorns designed for the millimeter-wave band of approximately 80~GHz to 170~GHz (i.e. a typical range for the 90/150~GHz CMB bands). 
We then present a full 150~mm diameter feedhorn array designed to operate in the submillimeter band centered at 350~GHz (330~GHz to 370~GHz).
This array is a duplicate of those produced via the traditional DRIE process for the CCAT 350~GHz instrument module of Prime-Cam\cite{vavagiakis2018ccatprime}.  
Together, these arrays present a successful demonstration of laser micromachining as a future solution in silicon-platelet feedhorn production with potential advantages in fabrication speed, throughput and cost. 
Sec.~\ref{sec:fab} presents the designs, fabrication and metrology of the produced feedhorn arrays.  
Sec.~\ref{sec:meas} presents beam and throughput measurements of multiple feedhorns in each array, comparing room-temperature optical measurements of both laser and DRIE processed feedhorns to simulation.
Finally, Sec.~\ref{sec:future} explores the future of laser-micromachining of silicon feedhorns, including the potential for sidewall control and the use of thicker (and thus fewer) wafers, which could significantly simplify production.  

\section{DESIGN AND FABRICATION}
\label{sec:fab} 

We use existing platelet feedhorn designs for these demonstrations as to provide side-by-side comparisons to traditional DRIE-etched platelet feedhorns that have already been produced for various ongoing experiments.  We first demonstrated laser-micromachined feedhorns with the 2-inch 90/150~GHz feedhorn array described in Sec~\ref{sec:mf_fab}. This was followed by the high-frequency and large-scale array for operation in the 350~GHz band as described in Sec.~\ref{sec:350_fab}.  

With the exception of the laser micromachining, these feedhorn arrays were constructed using the same well-established\cite{Britton2010,nibarger201284,simon2016,simon2018feedhorn}
methods used for DRIE-etched platelets that are recently detailed in Austermann et al 2026\cite{austermann26}.
In short, a number of platelets 
are microfabricated with unique hole sizes that, when stacked, produce an array of the desired feedhorn profile(s) within the array footprint.  
These platelets are cleaned\cite{nibarger201284} and each is metalized with a $\sim$1~$\mu$m Ti/Cu seed layer that is sputter deposited from each side of the wafer.  
These platelets are then precision aligned against multiple granite reference blocks that have sub-micron surface flatness.  
Two edges of the platelet stack are simultaneously aligned against the blocks by tapping individual platelets while monitoring under a microscope. Additional reference blocks maintain parallelism of the top and bottom surfaces to minimize racking (tilting) of the stack.  
Once aligned, temporary screws are tightened across the array to fix the alignment while small strips of epoxy are applied to the edges of the stack to set the alignment.  
The full array is then electroplated with a target of $\sim$3~$\mu$m of gap-filling Cu followed by $\sim$3~$\mu$m of electroplated Au.  These thick electroplated layers act to create a continuous conductor inside the feedhorns and prevent layer-to-layer separation.  
The temporary screws are removed after gold plating and the feedhorn array is considered complete. 

\subsection{Laser Micromachining and Debris Removal}
\label{sec:debris}

   \begin{figure} [ht]
   \begin{center}
   \begin{tabular}{c} 
   \includegraphics[width=16.7cm]{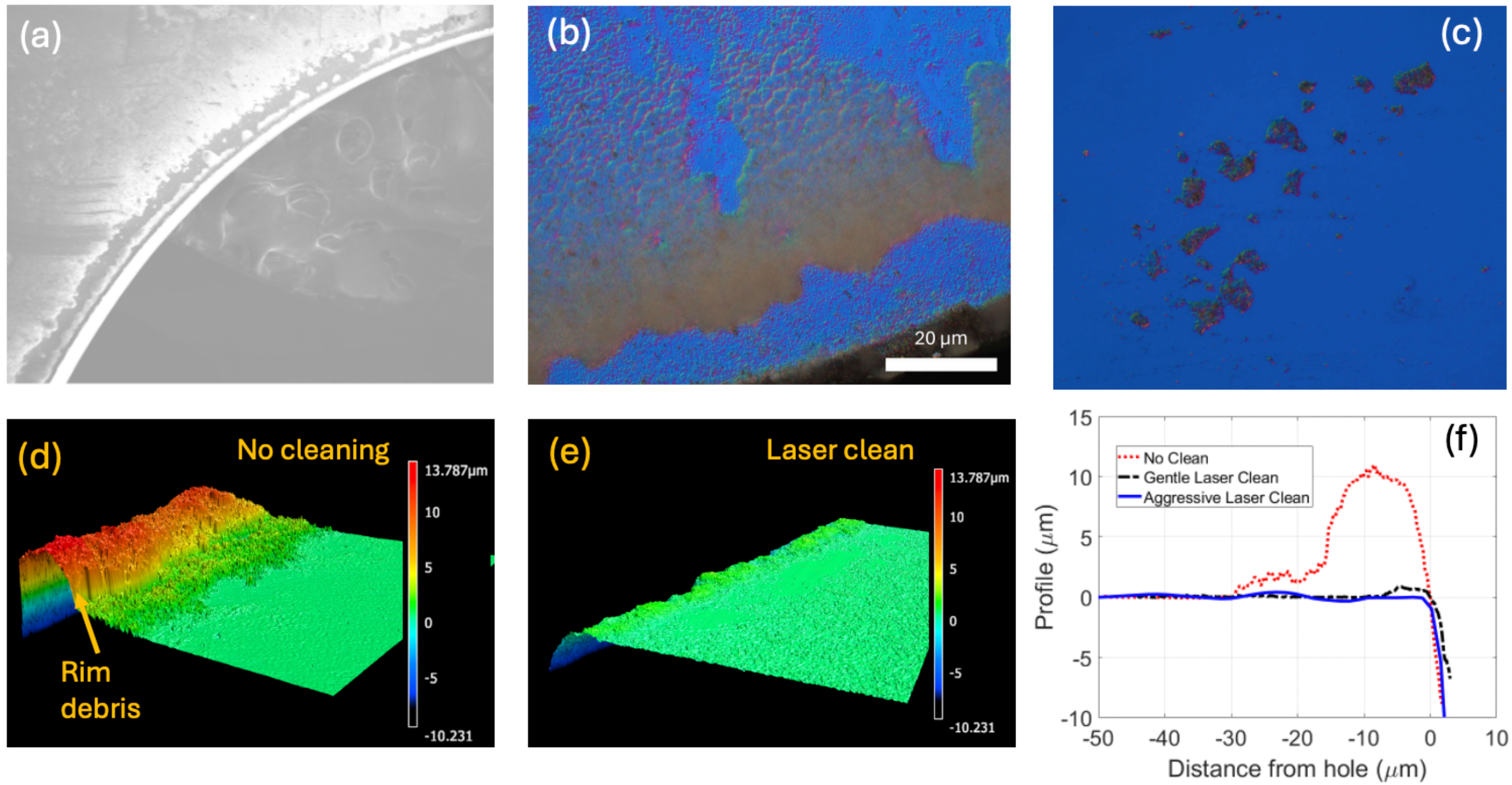}
   \end{tabular}
   \end{center}
   \caption[example] 
   { \label{fig:debris} 
Micrographs and profilometry of debris deposits before and after laser cleaning.  Top Row: (a) micrograph of a laser-micromachined hole and associated debris concentrated near the rim; (b) differential interference contrast (DIC) micrograph of the debris near the hole rim; (c) DIC micrograph of typical $\sim$1~$\mu$m high field deposits far from hole edge. 
Bottom Row: (d) 3D profilometry near a hole rim without laser cleaning; (e) 3D profilometry with laser cleaning; (f) 2D height profile of debris near the hole edge with and without various forms of laser cleaning. Laser cleaning parameters that resulted in a more gentle laser cleaning with less impact on the original silicon surface left some residual debris closest to the hole edge. More aggressive cleaning removed all debris and was the method used for all production wafers.  This process can lightly etch the base silicon, leaving a sub-micron texturing of the silicon surface. 
}
   \end{figure} 

The laser-micromachined platelets were produced at General Atomics using an XY galvanometer laser scanner processing system.  
Processing rates are a function of the total area and depth of micromachining required.  
To increase production rates, only an annulus of each hole's outer radius is cut, dropping out the hole center.   
A similar approach is used for DRIE processing where annuli are also used to minimize etch area for improved process control that can produce straighter sidewalls. 
Laser platelet production of the 150~mm wafers for the 350~GHz array (Sec.~\ref{sec:350_fab}) took a few weeks of relatively consistent effort for this demonstration. 
This is comparable to the time frame typically required to produce a similar number of platelets using the traditional DRIE process under normal workflows of a research level cleanroom.  
While this demonstration did not increase production rates, we note that laser micromachining is readily scalable in ways that would not be practical for cleanroom-based lithographic DRIE production\footnote{It is also possible to scale production with additional DRIE systems, lithographic equipment, and autoloaders to increase throughput. However, such equipment typically comes at a high cost, requires valuable and limited cleanroom space, and often necessitates more human oversight than the laser alternative.}.
The laser system used here was not optimized for production rate, but we expect it to be straightforward to increase fabrication speeds with an upgraded laser, a faster XY stage, and/or multiple lasers/stages.


During the laser micromachining process, some of the ablated silicon and its byproducts are redeposited on the bare silicon surface. 
In our process, we found debris could be $\gtrsim10$~$\mu$m high near the rim of the micromachined features as well as smaller $\sim$1~$\mu$m deposits further afield as described in Fig.~\ref{fig:debris}.  
We find that the debris is not easily removed using simple and common solvents (e.g. isopropyl, acetone) and/or abraded with a soft material (e.g. cotton swab or cloth).  While the debris can be removed through scraping with a hard material and sharp edge, such a process easily damages the silicon surface and risks catastrophic fracturing of the wafer. 

   \begin{figure} [ht]
   \begin{center}
   \begin{tabular}{c} 
   \includegraphics[width=16.7cm]{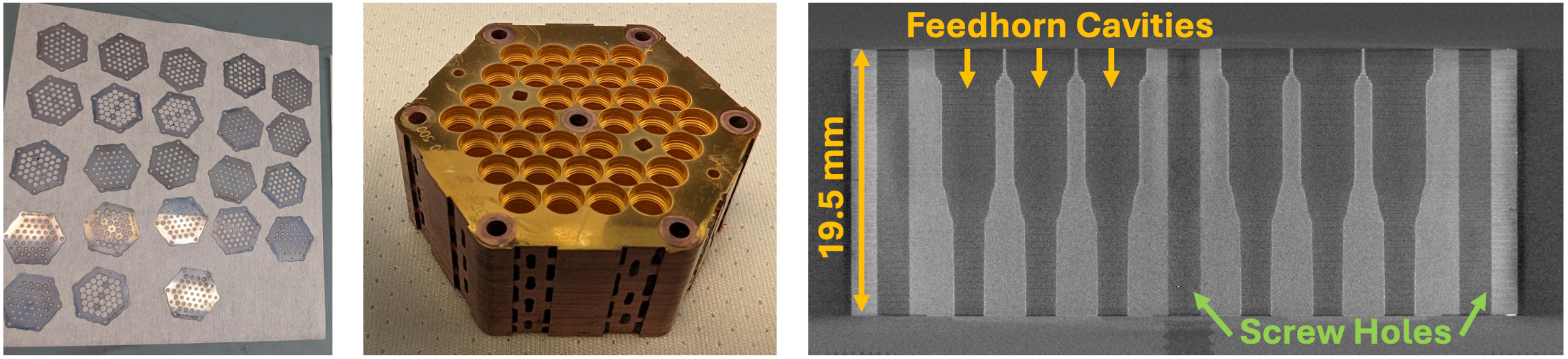}
   \end{tabular}
   \end{center}
   \caption[example] 
   { \label{fig:mf_fab} 
The 90/150~GHz silicon-platelet feedhorn array produced using laser micromachining.
(Left:) 23 of the 47 component platelets after laser micromachining and before Ti/Cu seed layer deposition.
(Center:) The completed 50~mm outer diameter array of 34 feedhorns.
(Right:) Example X-ray tomography of the completed array. 
Although full metrological analysis has not yet been completed, spot checks of dimensions show the achieved feedhorn profiles are consistent with design. 
Full 3D tomography allows for the construction of `digital twin' models of the realized feedhorn array.   
}
   \end{figure} 

Debris left on the surface can cause two primary problems for the silicon feedhorn construction process.  First, the tallest features would cause platelets to stack unevenly with potential gaps between layers.  The tallest features could also change the total depth of each step in the stack, changing the overall feedhorn profile. 
Second, we find that the sputtered Ti/Cu seed layer does not adhere well to the deposited material.  
Adhesion was tested using the application and removal of polyimide tape.  The tape easily removed the seed layer in areas of heavy debris, but consistently could not remove the seed layer in areas that were free from debris before metalization. 
This is the same tape test used in traditional DRIE processing and has proven valuable to detect poor adhesion, typically due to insufficiently clean wafers at the time of metalization.
Delamination could interrupt the otherwise continuously conducting feedhorn walls and leave dangling metal flakes in the feedhorn volume, resulting in increased loss and reflection within the feedhorn.

Deposits from laser ablation are a well-known problem\cite{rihakova2015laser} and there exists extensive literature on addressing the issue.  
Various techniques have been developed to significantly reduce deposits through fine-tuning of laser parameters, use of ultrashort-pulsed lasers\cite{gaudiuso2021laser,metzner2022ablation}, and ablating in a reactive gas\cite{haustrup2011laser} or liquid\cite{ren2005laser,choo2004micromachining} environment. However, these techniques are unlikely to completely eliminate all deposits. 
A protective layer can be applied to the surface before laser ablation and later removed\cite{simon2018feedhorn}. 
However, the laser kerf can remove the protective layer just beyond the feature's edge, resulting in a ring of exposed silicon surface that can receive deposits\cite{hazra2020thermal}.
For a surface effectively devoid of debris, an additional post-micromachining cleaning step is often required, e.g. a KOH or HNO$_3$/HF etching, which will also affect the native silicon surface.

While incorporating one or more of these techniques could prove sufficient for producing suitable platelets, we note that they all potentially add significant processing steps that may require specialized processing tools and/or the controls typical of the cleanroom environment we are trying to avoid.
The primary motivation of this work is to find a streamlined and scalable alternative to the established DRIE-based fabrication process; therefore, 
we sought a solution that could be fully automated and integrated into the fabrication process with little to no additional equipment or labor required. 
To accomplish this, we incorporated a `laser cleaning'\cite{zhou2023_lasercleaning} step to the process. 
Laser cleaning takes advantage of the difference between laser energy deposition in the debris that adheres to the surface of the silicon wafer, and the silicon wafer itself. The laser ablation process is adjusted so that the silicon substrate does not absorb enough energy to cause significant ablation, whereas the debris tends to absorb more energy and hence is ablated preferentially. Various modalities of this process are available, including liquid-assisted and precleaning to remove loose debris.\cite{zhou2023_lasercleaning}.
By fine-tuning the laser and scanning settings in a process developed by General Atomics, the deposited debris is ablated away with minimal impact on the original silicon surface.
Optical profilometry of the surface with and without the laser cleaning step is described in Fig.~\ref{fig:debris}.
The exposed silicon surface can become lightly textured through the process, typically at scales smaller than the $\sim$1~$\mu$m seed metalization.
We speculate that the texturing could enhance adhesion compared to a polished surface\cite{xiao2018adhesion}. 
The cleaning step uses the same laser system used for micromachining and can be fully automated into the fabrication process with no additional labor, equipment, or setup.  
Laser cleaning was performed on all parts used in both arrays presented in this work and no delamination of metalization layers or other deleterious effects have been observed.  

\subsection{The 90/150~GHz Demonstration Array}
\label{sec:mf_fab}

The first array built uses a modified version of the platelet feedhorn profile designed and optimized for Advanced ACTPol\cite{simon2016} (AdvACT).  
Like AdvACT, it is constructed from a combination of 250~$\mu$m and 500~$\mu$m thick silicon platelets for a total feedhorn length of 19.5~mm. 
The stackup has been simplified to 47 total platelets, as opposed to the 53 used in AdvACT, by merging pairs of 250~$\mu$m thick platelets into single 500~$\mu$m platelets where simulations show doing so has negligible effect on the feedhorn beam shape or performance.  This is the same simplified profile used in the first module of AliCPT-1\cite{salatino2020design}.  

The individual platelets are micromachined as 50~mm outer diameter hexagons, each with 34 holes for the feedhorns and additional holes for screws and alignment pins. 
The individual platelets were laser micromachined from 150~mm silicon wafers, yielding 7 platelets per wafer. 
Example platelets and the fully realized feedhorn array are shown in Fig.~\ref{fig:mf_fab}. 
Each platelet received laser cleaning at the end of the micromachining process.  
From here, the feedhorn platelets were cleaned, seed plated, stacked, aligned, and electroplated as described previously.  

Optical and/or x-ray metrology (Fig.~\ref{fig:mf_fab}) was performed on individual platelets and the full array at various points in the fabrication process.
This includes spot checks of hole sizes and positions, which were generally found to be within a few $\mu$m of design -- well within the target $\sim$25~$\mu$m tolerance typical for these frequencies. 
The achieved dimensions of the waveguide are often the most critical geometry of the feedhorn as it can be used as a high-pass filter that defines the lower edge of a detector's bandpass. 
Therefore, more comprehensive optical metrology was done on the waveguide exit apertures of the completed array and is summarized in Tab.~\ref{tab:metrology}. 
These measurements are after copper and gold electroplating of the array, which has itself been shown to have a thickness uncertainty of $\sim$3~$\mu$m relative to design\cite{austermann26}. 
Overall, the achieved geometries are well matched to the design with high uniformity and surpass the $\sim$20~$\mu$m tolerances one might expect from precision direct machining of metal feedhorns.  
Although within typical tolerance requirements, the amount of non-circularity is high relative to the other parameter accuracies and is a likely target for improvement through further investigation and process development.

\begin{table}[t]
\caption{Optical metrology measurements of the waveguide exit apertures of all feedhorns of the listed arrays. Positional and diameter accuracies are given as the averaged difference from design. 
Plus/minus values are the 1 standard deviation of the measured population.
Achieved diameters of the 90/150~GHz part include uncertainty in the electroplating thickness, which is expected to be relatively uniform across the array. 
}
\label{tab:metrology}
\begin{center}       
\begin{tabular}{|c|c|c|c|c|} 
\hline
\rule[-1ex]{0pt}{3.5ex} Measured Part Type & X Positional & Y Positional & Diameter  & Circularity  \\
\rule[-1ex]{0pt}{3.5ex}   & Accuracy [$\mu$m] & Accuracy [$\mu$m] &  Accuracy [$\mu$m] &  [$\mu$m] \\
\hline
\rule[-1ex]{0pt}{3.5ex}  90/150 Demo Waveguide Aperture & 3.1 $\pm$ 4.4 & 3.3 $\pm$ 6.2 & 3.6 $\pm$ 3.3 & 7.2 $\pm$ 2.2 \\
\hline
\rule[-1ex]{0pt}{3.5ex}  350~GHz Platelets (prototypes) & -0.8 $\pm$ 6.5 & -0.6 $\pm$ 9.9 & 0.1 $\pm$ 5.7 & 5.7 $\pm$ 3.9 \\
\hline
\end{tabular}
\end{center}
\end{table}

\subsection{The 350~GHz CCAT Feedhorn Array}
\label{sec:350_fab}

The 350~GHz feedhorn array shares the same design as the CCAT arrays produced with the traditional DRIE process\cite{austermann26} and is one of three total arrays deployed in the 350~GHz module of the Prime-Cam instrument\cite{vavagiakis2018ccatprime}.
The feedhorn is constructed from 33 individual platelets and uses a profile that has been optimized for performance in the atmospheric transmission window of 330~GHz to 370~GHz. 
Like the 90/150~GHz feedhorns, the array is constructed from a combination of 250 and 500~$\mu$m thick wafers, with the smaller step size being required where the profile
is rapidly changing in diameter and the small step size better approximates a smooth-walled feedhorn. 
The feedhorns are hexagonally packed with a 2.75~mm pitch for a total of 1,707 feedhorns.
The sky-side aperture is 2.65~mm, leaving 100~$\mu$m of interstitial silicon between apertures for structural integrity. The feedhorns and integrated waveguide are 10.5~mm in total length.

   \begin{figure} [t]
   \begin{center}
   \begin{tabular}{c} 
   \includegraphics[width=16.7cm]{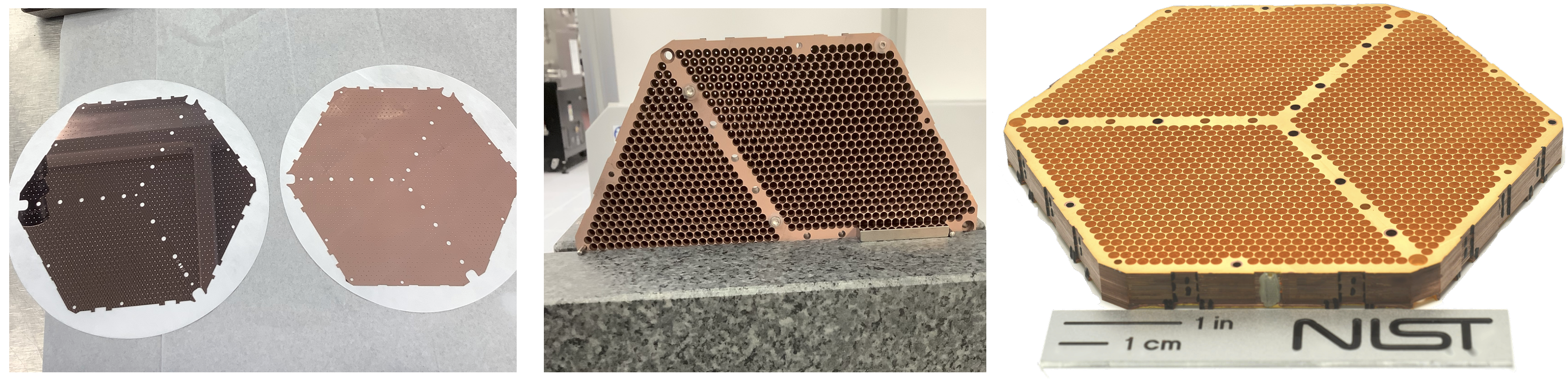}
   \end{tabular}
   \end{center}
   \caption[example] 
   { \label{fig:350_fab} 
The laser-micromachined 350~GHz feedhorn array. (Left:) Two of the individual platelets after seed plating. 
(Center:) The fully stacked array during the alignment process -- seen here between two precision granite blocks.
(Right:) The completed array after electroplating.}
   \end{figure} 

The timeline for integration with detectors and deployment to the telescope did not allow for full-scale metrology on input parts or the completed 350~GHz array.
However, wafer-scale metrology was performed on prototype, non-metalized platelets and is included in Tab.~\ref{tab:metrology}. 
Multiple production platelets and the full array were also spot-checked and found to be generally consistent with the hole accuracy and precision of Tab.~\ref{tab:metrology}.   
The micromachined holes were found to have sidewall tapers between 2.0~deg and 3.0~deg, and were consistent across the various hole sizes measured.
This results in the top of the hole (relative to the laser) being larger than the bottom.  
This is consistent with the taper seen in most previously produced DRIE platelets\cite{simon2016}, although $\sim0.5$deg sidewalls were recently demonstrated\cite{austermann26}.  
Average tapers of $\sim$2.5~deg are within tolerance of the designs and are further mitigated by fabricating and stacking the platelets such that the larger side of the hole faces the larger sky-side aperture, i.e. the larger side faces the growing profile diameter and often better approximates a smooth-walled profile than a purely vertical taper.
Furthermore, we reduced the target hole size such that the average hole diameter through the wafer is matched to design. 

Four 500~$\mu$m thick wafers are used to form a 2~mm long waveguide at the detector end of the feedhorn. 
The waveguide is designed with a diameter of 0.532~mm to provide a high-pass cutoff of $\sim$330~GHz and critically sets the low edge of the CCAT 350~GHz band.  
Since the production timeline did not allow for full metrology of potential laser-micromachined waveguide platelets, these four wafers were produced using the traditional DRIE process. 
This choice was made to minimize risk of possible unknown systematic differences in hole geometries between the DRIE and 
laser-micromachined arrays that could otherwise cause differences in bandpass shapes and detector coupling between the three arrays of the CCAT 350~GHz focal plane. 
All other platelets (29 in total) were produced using laser micromachining.  
Like the other two CCAT 350~GHz feedhorn arrays\cite{austermann26}, this array underwent two rounds of electroplating to further tune the cutoff frequency.

   \begin{figure} [t]
   \begin{center}
   \begin{tabular}{c} 
   \includegraphics[width=16cm]{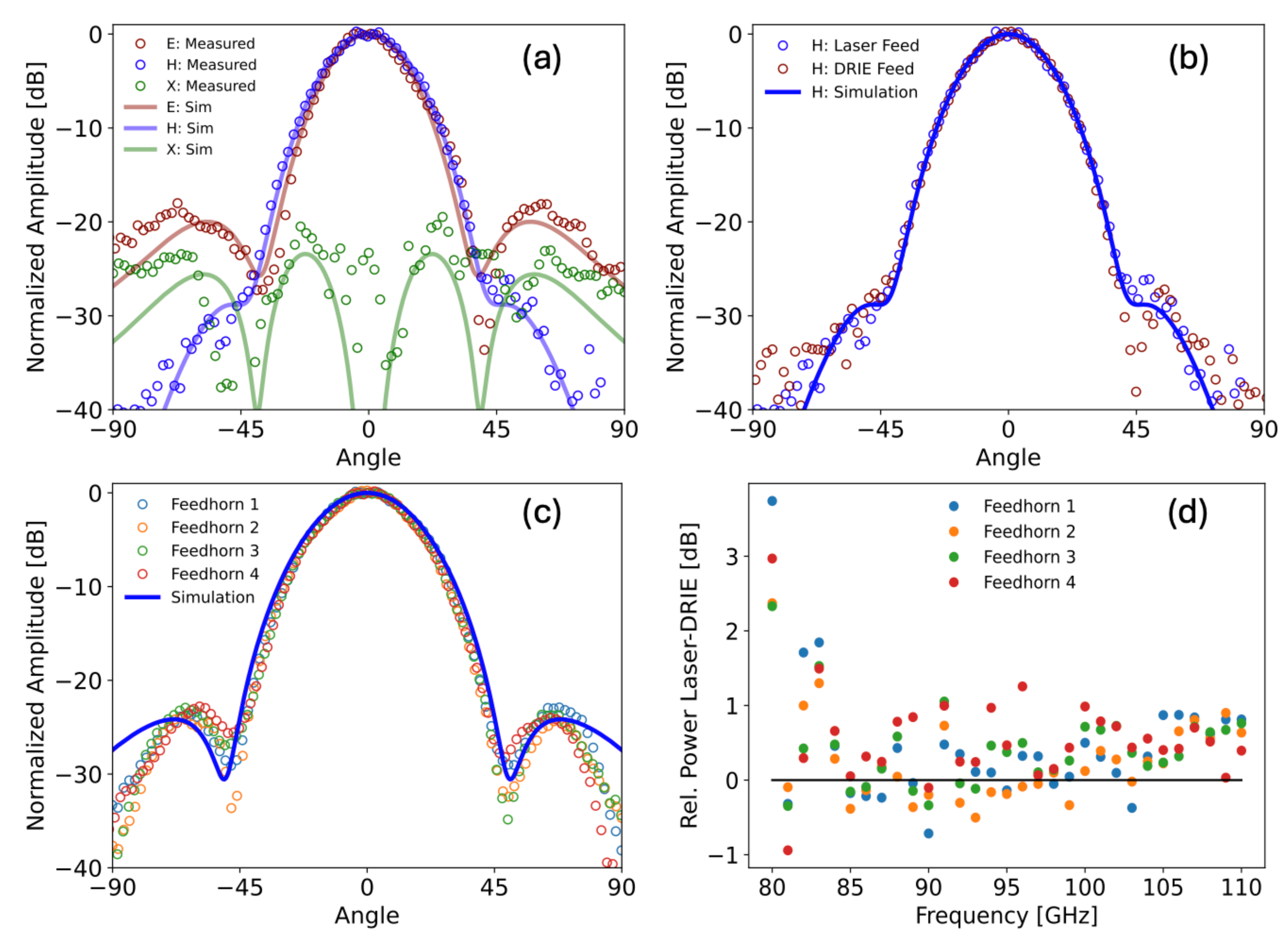}
   \end{tabular}
   \end{center}
   \caption[example] 
   { \label{fig:mf_meas} 
VNA beam and relative power measurements of the 90/150~GHz feedhorn array.
(a) E-, H-, and D-plane (cross-polarization, X) beam measurements of a single feedhorn at 155~GHz (data points) compared to simulation (solid lines).  Some reflections in the measurement setup cause excess variation at beam center. 
(b) A direct comparison of 155~GHz H-plane measurements of a laser-micromachined feedhorn to one produced using the traditional DRIE process.
(c) Array uniformity: H-plane measurements at 110~GHz for four feedhorns at different locations across the array.
Additional absorbing material around reflective areas of the measurement setup significantly reduced reflections at beam center compared to the earlier measurements of plots a \& b.  The absorbing material is likely responsible for the excess attenuation at the highest angles.
(d) Transmitted power of the same four feedhorns relative to the power transmitted in the DRIE feedhorn. High variation at the lowest frequencies is likely due to slightly different diameter waveguides from variation in electroplating thickness as well as any residual bias between DRIE and laser processing. 
}
   \end{figure}

\section{MEASUREMENTS AND CHARACTERIZATION}
\label{sec:meas} 

\begin{figure} [t]
   \begin{center}
   \begin{tabular}{c} 
   \includegraphics[width=16cm]{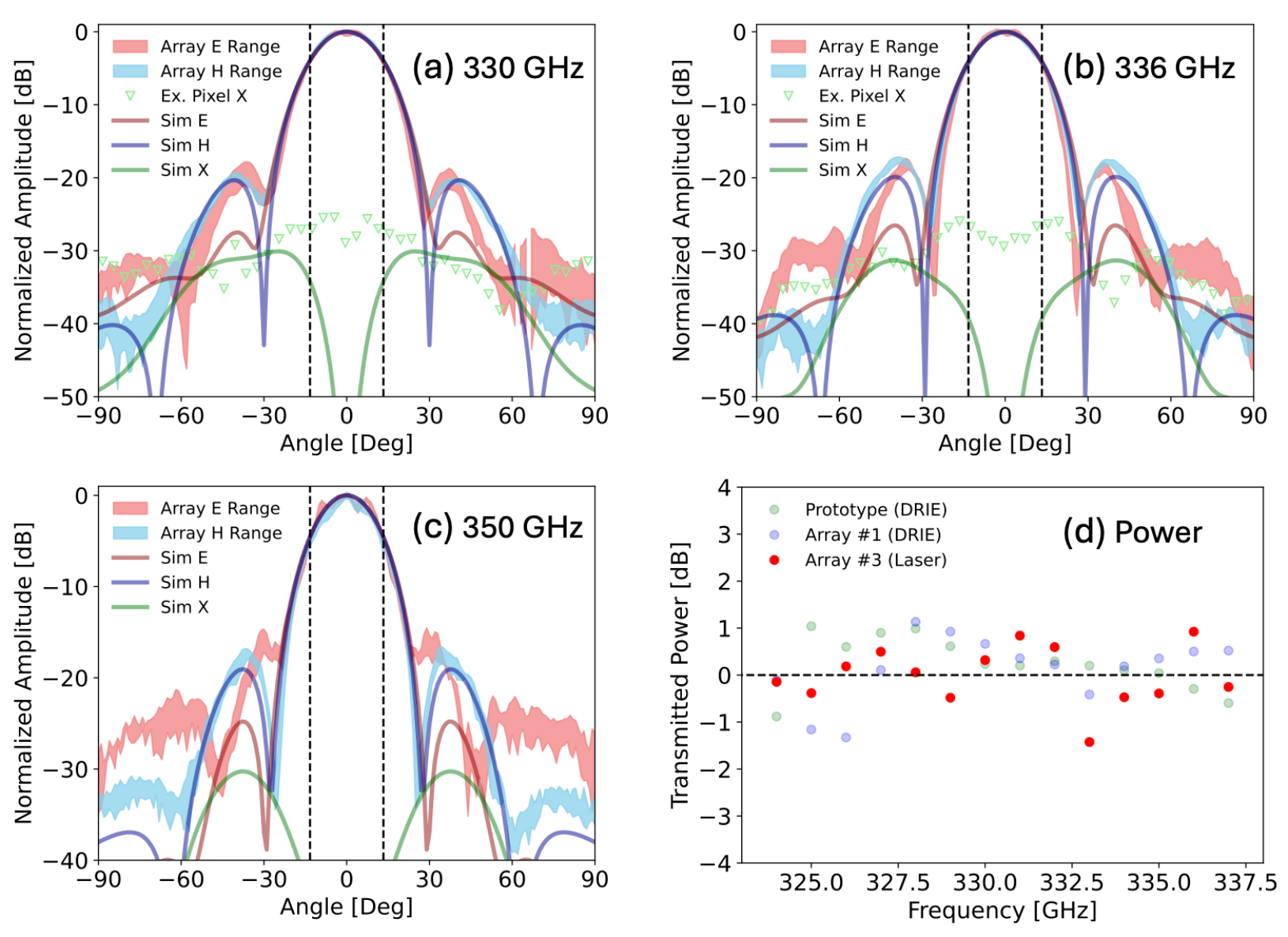}
   \end{tabular}
   \end{center}
   \caption[example] 
   { \label{fig:350_meas} 
VNA beam and relative power measurements of the 350~GHz feedhorn array. 
Comparable DRIE horn measurements are published elsewhere\cite{austermann26}, where they are found to be well matched to simulation.
(a) E- and H-plane beam measurements at 330~GHz for six different feedhorns distributed across the array ($\pm$1 standard deviation shaded region) compared to simulation (solid lines).  Measurements are made using a WR3.4 waveguide setup.
D-plane cross-polarization measurements were made for one horn (data points).
Vertical dashed lines represent the opening angle to the CCAT 350~GHz module optical cold stop and represents the part of the beam that reaches the sky.
(b) Same as (a) but at 336~GHz.
(c) Same as a/b but at 350~GHz using a WR2.2 waveguide setup that operates at a lower absolute power (resulting in lower signal-to-noise measurements) and exhibits more system related systematics (reflections and resonances unrelated to the feedhorn).  
We did not have access to WR2.2 waveguide twists to allow for D-plane measurements of cross-polarization at 350~GHz.
(d) Average transmitted power of laser feedhorns and other DRIE horns relative to the power transmitted by a DRIE produced CCAT feedhorn array of the same design. 
Fewer feedhorns received power measurements; each data set consists an average of two feedhorns while the reference data set (Array \#2) is the average of four feedhorns.
}
   \end{figure}

We measure the beam profiles of multiple feedhorns across each of the arrays using a room-temperature 
millimeter-wave vector network analyzer (VNA) 
setup similar to that described in previous work\cite{simon2016,austermann26,stevenson2026}.   
The feedhorn under test is used in broadcast mode as the transmitter while a receiver is placed in the far-field and swept in angle.  
This is repeated for various polarization planes (E, H, D) using waveguide twists between the feedhorns and the source/receiver. 
Commercial diagonal feedhorns are used on the receiver for all measurements except for the 350~GHz band cross-polarization measurements where a prototype CCAT 350~GHz feedhorn was used for its low demonstrated cross-polarization\cite{austermann26}.  
Measurements of the 90/150~GHz feedhorns and the 350~GHz feedhorns are shown in Figs.~\ref{fig:mf_meas} \& \ref{fig:350_meas}, respectively.

Overall, the beams produced by both feedhorn arrays are well matched to simulation, as well as to DRIE-produced feedhorns.  
However, we do measure enhanced side lobes in the E-plane beam of the laser-micromachined 350~GHz feedhorns relative to both simulations and the DRIE feedhorns\cite{austermann26}.  
These excess side lobe features are seen at all frequencies in the band and are comparable in size to the designed and measured H-plane side lobes.  
We are currently investigating the source(s) of these higher side lobes, which we speculate may be due to a combination of systematically over-sized holes, hole ellipticity, and/or misalignment in the feedhorn stack. 
Fortunately, these side lobes are still relatively low (peaking at $\sim$~-20~dB) and well outside the part of the primary beam that projects out of the Prime-Cam optics (dashed lines in Fig.~\ref{fig:350_meas}). 
In Prime-Cam, the side lobes are fully terminated on an optical cold stop and are expected to have a negligible effect on the on-sky performance. 

The feedhorn beams show high levels of uniformity across the arrays with scatter comparable to that seen in the DRIE produced 350~GHz arrays.  
As described in Austermann et al.\cite{austermann26}, alignment-related systematics in the measurement result in a similar scale of scatter; therefore, the observed spread in beam measurements provides an upper limit to the inherent variation of beam properties across the array.

We also use the VNA measurements as a rough check for excess loss in the feedhorns. 
Although gain drifts and other systematics in the measurement system limit power estimates to approximately $\pm$1~dB, the measurements in Figs.~\ref{fig:mf_meas}d \& \ref{fig:350_meas}d are consistent with there being no excess loss in the laser-micromachined feedhorns compared to those produced with traditional DRIE.

\section{OUTLOOK AND FUTURE WORK}
\label{sec:future}  

   \begin{figure} [t]
   \begin{center}
   \begin{tabular}{c} 
   \includegraphics[width=16.7cm]{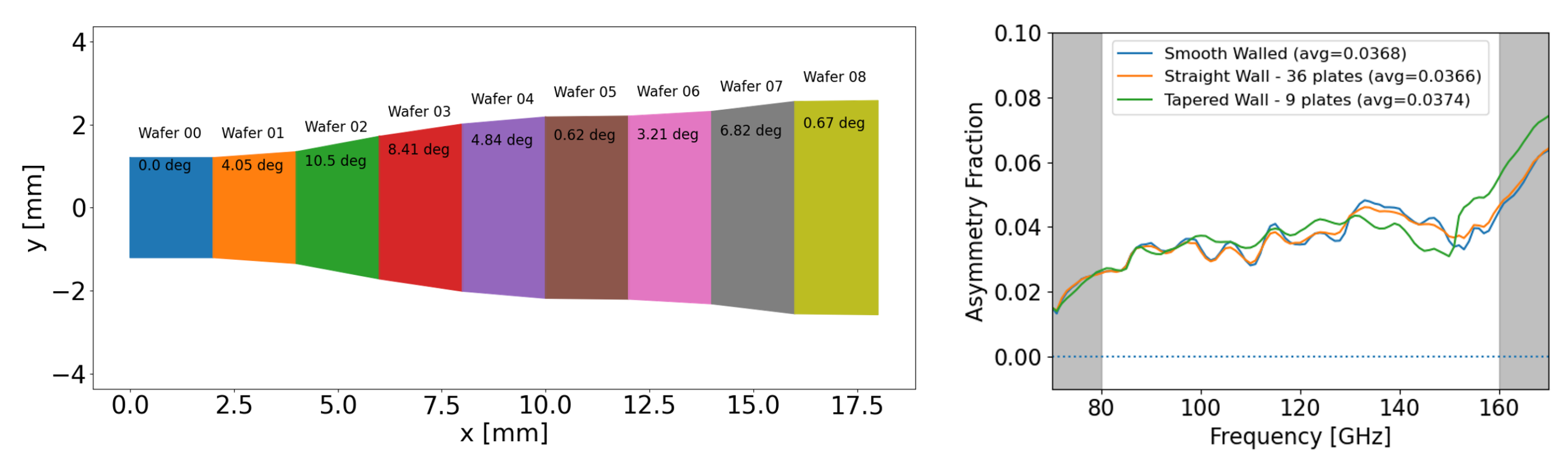}
   \end{tabular}
   \end{center}
   \caption[example] 
   { \label{fig:future}
   An example feedhorn design made from a small number of thick platelets with controlled-taper holes.  
(Left:) An interpolated version of the Simons Observatory 90/150~GHz feedhorn design consisting of just nine 2~mm thick platelets. 
(Right:) Beam asymmetry, comparing the original smooth-walled Simons Observatory design to a traditional straight-walled platelet design of 36 platelets and the sloped-walled 9 platelet version on the left. 
The y-axis represents the difference in power within the E- and H-plane beams, as defined in Eq.~\ref{eq:beam_asymmetry}, within a 13~deg half-width opening angle to a hypothetical cold aperture stop. Since this is only calculated for the central beam, this means the asymmetry values are for differences in the central beam and unrelated to side lobes.
 }
   \end{figure}

This work demonstrates that laser-micromachined silicon-platelet feedhorns are a viable option for high-precision work at millimeter/submillimeter wavelengths.
Furthermore, we believe these processes are readily scaled to the large production volumes required by potential future experiments.
Fabrication of platelets is the most significant component of silicon-platelet feedhorn production in terms of labor, costs, and bottlenecks.  
Moving from DRIE processing to laser micromachining could help alleviate all of these issues.

The production process still relies on sputter-deposited seed metalization of each platelet, which is typically done in a cleanroom. This step represents the next largest component of the total effort and cost of production. It could also prove to be the next bottleneck in production scalability. Therefore, reducing the total number of platelets in the design would not only streamline the whole process, but could also significantly reduce cost and increase throughput. 

Platelet feedhorn profiles are already optimized to be as short as possible without a reduction in the desired performance metrics. 
Past designs have typically used a maximum wafer thickness of 500~$\mu$m. 
Both DRIE and laser etching could move to thicker wafers.  
For the traditional straight sidewall designs that approximate a smooth-walled profile with sub-wavelength steps, the maximum wafer thickness is wavelength dependent\cite{austermann26}.  
At the wavelengths discussed here (i.e. $\gtrsim$~2~mm), wafers significantly thicker than 500~$\mu$m could only be used in the straightest parts of the feedhorn profile without impacting performance.  For the types of profiles discussed here, this would likely only reduce the total number of platelets by 10\% to 30\%. 

Instead, we propose using thick wafers where holes are micromachined with intentional and controlled sidewall tapers.
In Fig.~\ref{fig:future}, we outline an example where we approximate the Simons Observatory 90/150~GHz feedhorn profile\cite{simon2018feedhorn,mccarrick2021_SO_UMM} using tapered sidewall platelets.
This design reduces the number of wafers by a factor of 4x compared to an approximation of the same horn profile using straight sidewall platelets. 
We find that this design has comparable performance to both the original smooth-walled design and a 36-layer straight-walled design. 
We compare the simulated beam symmetry of these three manufacturing options in Fig.~\ref{fig:future}, defined here as power difference between the 2-dimensional E and H beams relative to the total power within an aperture stop:

\begin{equation} \label{eq:beam_asymmetry}
\rm{Beam~Asymmetry} = \frac{\int_{0}^{\theta_{\rm{stop}}} (H^2 - E^2)~sin(\theta)~d\theta}
{\int_{0}^{\theta_{\rm{stop}}} (H^2 + E^2)~sin(\theta)~d\theta},
\end{equation}
where E and H are the amplitudes of the E-plane and H-plane beams (respectively), $\theta$ is the radial coordinate of the beam, and $\theta_{\rm{stop}}$ is the radial angle to the edge of the aperture stop.

Simulations also show no significant difference, i.e. $<$~0.1\% in power, in spillover efficiency (beam coupling\cite{simon2016}), cross-polarization, and reflection within a hypothetical 13~deg half-width opening angle to a cold aperture stop. 
We also note that the sloped etch design is merely an interpolation of a previously optimized profile
and it may be possible that a dedicated optimization of a sloped etch design could further improve performance. 
Building full arrays of feedhorns from such a small number of wafers would significantly reduce both manufacturing overheads and seed plating costs, even if the total required silicon micromachining time is similar to the traditional approach.
Controlled tapered sidewalls can be produced in silicon with either DRIE\cite{roxhed2007tapered} or with laser micromachining through angle control of the laser or positional changes during ablation. 
However, laser micromachining remains a preferred solution due to its advantages in scalability and reduced cleanroom time.

\acknowledgments 
 
The laser micromachining program was funded in part by the General Atomics Internal Research \& Development (IRD) fund.  We thank the CCAT collaboration for the opportunity to build and field a laser-micromachined feedhorn array for their 350~GHz imaging module. Author Sara M. Simon recognizes that this manuscript has been authored by FermiForward Discovery Group, LLC under Contract No. 89243024CSC000002 with the U.S. Department of Energy, Office of Science, Office of High Energy Physics. 

\bibliography{journals,jh,mkidbib,citations}
\bibliographystyle{spiebib} 

\end{document}